\documentclass[useAMS,usenatbib]{mn2e}
\usepackage{graphicx,rotate}
\usepackage{natbib}
\usepackage{latexsym,amssymb,verbatim}

\title[Warm Cores around Regions of Low-Mass Star Formation]{Warm Cores around Regions of Low-Mass Star Formation}
\author[Zainab Awad, Serena Viti, Mark P. Collings, and David A. Williams]{Zainab Awad$^{1}$\thanks{E-mail:
zma@star.ucl.ac.uk} Serena Viti$^{1}$, Mark P.Collings$^{2}$, and David A. Williams$^{1}$ \\
$^{1}$Department of Physics and Astronomy, University College London, London WC1E 6BT, UK\\
$^{2}$School of Engineering and Physical Science, Heriot-Watt University, Edinburgh EH14 4AS}

\begin{document}

\date{Accepted 2010 May 24. Received 2010 May 24 ; in original form 2009 October 28}

\pagerange{\pageref{firstpage}--\pageref{lastpage}} \pubyear{2010}

\maketitle

\label{firstpage}

\begin{abstract}
Warm cores (or hot corinos) around low-mass protostellar objects show a
rich chemistry with strong spatial variations. This chemistry is generally
attributed to the sublimation of icy mantles on dust grains initiated by
the warming effect of the stellar radiation. We have used a model of the
chemistry in warm cores in which the sublimation process is based on
extensive laboratory data; these data indicate that sublimation from mixed
ices occurs in several well-defined temperature bands. We have determined
the position of these bands for the slow warming by a solar-mass star. The resulting
chemistry is dominated by the sublimation process and by subsequent
gas-phase reactions; strong spatial and temporal variations in certain
molecular species are found to occur, and our results are, in general, 
consistent with observational results for the well-studied source IRAS 16293-2422. The model used is
similar to one that describes the chemistry of hot cores. We infer that
the chemistry of both hot cores and warm cores may be described by
the same model (suitably adjusted for different physical parameters).

\end{abstract}

\begin{keywords}
Astrochemistry - Stars: low-mass, formation - ISM: abundances, molecules
\end{keywords}

\section{Introduction}
The early stages of low-mass star formation are observed
at millimetre and submillimetre wavelengths, where both the dust continuum
as well as molecular emission from the outer regions of the envelope can be
detected.  The dynamical aspect of low-mass star formation, with low
mass stars forming from dense and cold condensations called prestellar
cores, and evolving into Class 0 and I sources, is relatively well
defined. What is less well understood is how prestellar cores evolve chemically
into protostellar cores which then, eventually, develop
protoplanetary disks.  Nevertheless, it is now clear that, as in the
case of massive star formation, towards the end of the collapse phase
the gas and dust are dense and cold enough that (other than H$_2$ and He) 
material has frozen onto the
dust and that surface chemistry has taken place, leading to a rich gas
phase chemistry once the star heats the surrounding dust so that icy 
mantles returned to the gas phase. The warm
cores where the icy mantles sublime are often called `hot corinos'
(e.g. \citealt{cec04asp,cec07msl}) due to their chemical 
resemblance with hot cores.
Recent observations toward several warm cores around low-mass
solar-like protostars have shown that they are rich in large, complex
organic molecules (COMs) including methanol CH$_3$OH, methyl formate
HCOOCH$_3$, methyl cyanide CH$_3$CN, and more complex species such as
ethylene glycol (CH$_2$OH)$_2$ (e.g. \citealt{caz03,bot04,hol04,bis08}). 
For a comprehensive review on complex organic interstellar molecules 
we refer the reader to \citet*{her09}. 
In addition, they show some interesting deuterated species (e.g. \citealt{cec98a,caux05}), 
and long-chain unsaturated hydrocarbons and cyanopolyynes \citep{sak08}.
In fact, the envelope of Class 0 sources can be divided
chemically into two zones: the zone where all the ices sublime (dust temperature $\ge
100$ K), i.e. the warm core region, and the outer cold envelope (dust
temperature $\le 100$ K) (e.g. \citealt{cec05iau}). The chemistry in the outer envelope
is similar to that in pre-stellar cores, which are cold ($\sim 10$ K) and dense
( $\le 1\times10^{5}$ cm$^{-3}$) condensations believed to be
precursors of the Class 0 sources \citep{ber07}. Many molecules in the outer envelope are 
frozen onto the grain mantles, so that their abundances are often lower than those in
molecular clouds. In the inner regions, sublimation is clearly important. 

The first model of the thermal structure of a low-mass protostellar
envelope was developed by \citet{cec96}. In this model, the
authors considered the chemistry and resulting line emission within
the inside-out collapse model \citep{shu77} but down to much smaller scales of $\sim
10$ AU. In their model, the gas temperature in the innermost regions
of the envelope reaches $\sim 100$ K resulting in the sublimation of
H$_2$O from grain mantles. 

More recent models (e.g. \citealt{cec20b, mar04}) suggest that molecular emission 
(e.g. from H$_2$CO) is emitted by two components: a cold H$_2$CO-poor outer
envelope and a warm H$_2$CO-rich core. The model calculations show a
dramatic increase in the abundances of H$_2$CO, H$_2$O, and SiO in the
inner, warmer, and denser region of the envelope. This abrupt change
occurs when the dust temperature exceeds 100 K and grain mantles
sublime. 

Evidently, in situations where a
cool outer envelope and a warmer core exist, sublimation - and the subsequent gas-phase chemistry that
this causes - will be important. Also, the sublimation will be a time- and space-dependent event as the
central protostar warms up. Given the role of sublimation, it is important to take account of
recent laboratory studies of the sublimation of ices \citep{col03a,col04} which show that different
species enter the gas phase at different temperatures, and that the sublimation occurs not continuously
through a gradual warm-up but in several well-defined and narrow temperature bands. Such a process may
give rise to a rather different gas-phase chemistry than a process in which instantaneous sublimation of
all species in a mantle occurs, as found to be the case for hot cores \citep{viti04}. 

Our purpose in this paper is, therefore, to model the sublimation process, taking account of the
laboratory data, in the context of warm cores in low-mass star-forming regions. We wish to assess whether
the more complex sublimation mechanisms indicated by the laboratory studies significantly affect the
detectable chemistry in these objects, as compared to models in which the desorption is instantaneous for
all species. The model we use is developed from a similar model used for the study of hot cores in
regions of massive star formation \citep{viti04}. Further, if the two similar models can successfully
describe cores in both types of star-forming region, then this may be taken as support for the view that
star formation is carried out by essentially the same process, regardless of mass. We describe our model
in Section 2, the results are given in Section 3, and we make some conclusions in Section 4.

\section{The Model}
\label{sec:model}

In order to investigate the chemical evolution of low-mass star
forming regions, we have used a time-dependent gas-grain chemical
model adapted from the \citet{viti04} hot core model. In this work, our model 
follows the chemical evolution of a free-fall collapsing cloud in the 
process of forming a low-mass star, then explores the chemical evolution of the 
remnant of the warm core located in the vicinity of the newly formed star.
The core is represented by a uniform slab subdivided into 5 shells represented by 5 
depth points of increasing visual extinction from the edge of the core to its centre.
The results are obtained from a two-phase calculation: Phase I, the collapsing phase, 
in which diffuse, mainly atomic material of initial number density $\sim 400$ cm$^{-3}$ 
at a temperature of 10 K undergoes collapse following the so-called modified free-fall collapse
described in \citet {raw92} until the number density reaches the typical density of a warm core 
(treated as a free parameter). The initial elemental abundances relative to hydrogen that we
have adopted are listed in Table 
\ref{tab:initial} \citep{viti04}. During this time, gas-phase chemistry occurs and atoms and molecules are 
depleted on to grain surfaces and hydrogenate when possible. 
The depletion efficiency is determined by the fraction of the gas-phase material that is frozen 
on to the grains. This fraction is arranged by adjusting 
the grain surface area per unit volume, and assumes a sticking probability of unity for all species. 
The fraction of material on grains is then dependent on the product of the sticking probability and 
the amount of cross-section provided per unit volume by the adopted grain size distribution. 
Note that, as the chemistry is time-dependent, different species
form at different times, and as a consequence the material frozen out on the grains at any one time is
representative not of the whole gas but of selected species.
Apart from direct hydrogenation, the only other surface reactions we include are the formation of 
CH$_3$OH and H$_2$CO from CO and of CH$_3$CN from the reaction of methane, CH$_4$, with HCN, as it has been shown 
that gas phase reactions are not sufficient to form CH$_3$OH and CH$_3$CN (e.g. \citealt{tiel82,wat03,gar08}). 
\par 
In Phase II, the warming-up phase, we follow the chemistry of the remnant core, 
after the star is born, for $\sim$ 10$^7$ yrs, assuming a uniform density throughout the core. 
In this phase, the central star heats up the surrounding gas and dust, causing sublimation of the 
icy mantles.  In the present work, we simulate this heating effect in the same way as in \citet{viti04}. 
We assume that the presence of an infrared source in the centre of the core or in its vicinity
causes an increase in the gas and dust temperature. The temperature is a
function of the luminosity (and therefore age) of the protostar as in
\citet{viti04}, but here we adjust the luminosity/mass power law to account for lower
protostellar masses than in the hot core case. The temperature reaches its maximum 
(assumed to be 100 K) at $\sim10^{5}$ years which is the assumed (typical) age of Class 0 sources \citep{and93}.

During the warming-up phase (Phase II) of the core, the mantle species desorb in various
temperature bands. \citet{col03a,col04} reported results of Temperature Programme Desorption (TPD) 
experiments for a large 
number of molecules. They showed that molecular ices can be divided into five categories: 
(i) CO-like; (ii) H$_2$O-like; (iii) intermediate; (iv) reactive; 
(v) refractory. For each category one can estimate the fraction 
of a particular molecular species that is desorbed in various temperature bands. 
These bands are from 1) the pure species; 2) a monomolecular layer on H$_2$O ice;
3) desorption during the amorphous-to-crystalline H$_2$O ice conversion (the `volcano'
effect); 4) co-desorption when the H$_2$O ice desorbs; and 5)
desorption of the bare grain surface (see \citealt{col04} for
more details).

Based on the previously mentioned experimental results, \citet{col03b} 
constructed a rate model to explain desorption processes in
a phenomenological manner over relevant astronomical time scales.
By running this chemical kinetic simulation of
water-ice desorption, we calculated the different desorption
temperatures using a power-law temperature profile fitted to a
Sun-like star. \citet{viti04} used the observed luminosity function 
of \citet{mol20}, and correlated 
the effective temperature of the gas with the age of the accreting 
protostar through a simple power law. \citet*{nom04} solved the 
radiative transfer equation for a clump warmed by a central star assuming that 
the radial profile of the temperature is defined as the inverse square 
law $T(r) \propto r^{-1/2}$, which is comparable to the profile obtained 
by \citet{sch02} in which $T(r) \propto r^{-0.4}$.
In our model, the temperature evolves in both time and space. 
The temperature profile in our model, shown in Eq.(1), is 
derived by combining the approach of \citet{viti04} with that of \citet*{nom04}: 

\begin{equation}
T_{d}(t,d) = 10 + A (t)^{B} \times (d/R)^{-0.5} \quad $K$
\end{equation}

where $T_d(t,d)$ is the temperature profile of gas and dust
in the core surrounding the stellar object, $t$ is the evolutionary age 
of the collapsing core, $d$ is the distance from the core
centre, and $R$ is the core radius. $A$ and $B$
are two constants derived from the boundary conditions, in this case,
the temperature at $\sim$ 150 AU from the star at $t$ = 0 (T = 10 K) and 
$t$ = 10$^5$ yrs (T = 100 K). The latter condition corresponds to the temperature 
obtained by \citet{sch02}.
We find that the temperature profile of a typical warm core capable of reproducing 
with high accuracy the empirical assumptions we consider is given by 

\begin{equation}
T_{d}(t,d) = 10 + 0.1927 (t)^{0.5339} \times (d/R)^{-0.5} \quad $K$
\end{equation}
Eq. (2), was then inserted in the chemical kinetic program \citep{col03b} to
estimate the position of the different desorption bands. 
From this equation it can be seen that the temperature rises slower than in the case
of high mass stars \citep{viti04} affecting, as a consequence, the times at which the desorption of 
the different species occur. The desorption mechanisms, throughout the core, 
are time-  as well as space-dependent processes. The closer to the heating source
the earlier the desorption occurs. The volcano and co-desorptions will occur at slightly different
temperatures (and times) depending on the depth within the core but they are around 84 K 
(at $\sim$ 5$\times$10$^4$ yrs, for the inner shell) and 95 K (at $\sim$ 7$\times$10$^4$ yrs, 
for the inner shell) respectively. 
These results are in line with those previously calculated for hot cores where the 
desorption temperatures decrease as a function of slowing down of the heating rate 
so that the lower the stellar mass, the lower the volcano and co-desorption
temperatures (see Table 2 in \citealt{viti04}).

Our chemical network is a modified version of the UMIST database \citep{let20} with 127 gas phase species and 
42 mantle species interacting in 1871 chemical reactions. Photo-reactions were included, taking into 
account both the external interstellar radiation field and the internal cosmic ray induced UV field.  
Both direct and indirect ionisation by cosmic rays
were also included, using a cosmic ray ionisation rate, $\zeta$, of
$1.3 \times 10^{-17}$s$^{-1}$ \citep{lep92}. 
Observations of emission lines (e.g from H$_{2}$CO and H$_{2}$O) toward low-mass 
protostellar objects reveal that there is a gradient in both density and temperature. 
The dust temperature reaches 100 K in cores where the densities vary between 
$1 \times 10^{7}$cm$^{-3}$, such as the case of IRAS 16293-2422 (e.g 
\citealt{cec20a,cec20b,cec04asp}) and $2 \times 10^{8}$cm$^{-3}$ 
for NGC 1333-IRAS 4A, B \citep{mar04}, and their radii range from 200 AU down to 27 AU. 

In order to investigate the sensitivity of the chemistry to the physical parameters of
warm cores around low-mass stars, we ran a grid of  
chemical models where we varied
(i) the final density of the collapsing core and
(ii) the percentage of the accreted species on to grain surfaces.
A characteristic of our chemical model is that we simultaneously follow the chemistry
as a function of time and depth: the mantle evaporation is temperature-dependent and the 
temperature profile is dependent on both the distance from the core centre and time.
Table \ref{tab:initial} summarises the various physical parameters used in our model. 
The ice composition at the end of Phase I
will depend both on the final density of the collapsing core and the percentage of accreted 
species on the dust. Typical abundances (with respect to the total number of hydrogen nuclei) 
for high density and high depletion, for the most relevant species i.e H$_2$O, CO, H$_2$CO, 
CH$_4$, CH$_3$OH, are 4.0$\times$10$^{-4}$, 3.7$\times$10$^{-5}$, 3.6$\times$10$^{-9}$, 
1.4$\times$10$^{-4}$, and 3.8$\times$10$^{-7}$ respectively. 
The computed abundance of CH$_4$ in the ice is high and arises as a consequence of our 
assumption that hydrogenation (in this case, of carbon) is efficient. Detection by ISO of 
CH$_4$ ice show lower values \citep{vand04}. However, the ISO beam encompasses material 
outside the core and the ISO observations are of young stellar objects in which sublimation may 
have already reduced the CH$_4$ component. Hence, the detection may be under-estimating the CH$_4$ 
ice abundance within the pre-stellar core.

\begin{table}
   \centering
\caption{ Model initial physical parameters and elemental abundances relative to hydrogen.}
  \label{tab:initial}
    \leavevmode
    \begin{tabular}{ll} \hline
 \multicolumn{2}{c}{Physical parameters} \\ \hline \\
Core density $\dag$ & $1.0 \times 10^{7}$ - $2.0 \times 10^{8}$ cm$^{-3}$ \\
Core maximum temperature & 100 K \\
Core radius & 30 - 150 AU \\ 
Depletion percentage $\dag$& 85 - 100\% \\ \hline
\multicolumn{2}{c}{Initial elemental abundances} \\ \hline \\
   Carbon & $1.79 \times 10^{-4}$ \\
   Oxygen & $4.45 \times 10^{-4}$ \\
   Nitrogen & $8.52 \times 10^{-5}$ \\
   Sulphur & $1.43 \times 10^{-6}$ \\
   Helium & $7.50 \times 10^{-2}$ \\
   Magnesium & $5.12 \times 10^{-6}$ \\ \hline
\end{tabular}\\
\flushleft
$\dag$ This parameter varies only during the collapsing phase (Phase I).\\
\end{table}

\section{Results}
\label{sec:results}
\subsection{Chemical trends}
 Figure 1 shows how the computed fractional abundance (with respect to the total number of hydrogen
 nuclei) of selected species varies as a function of the radius of the core at around 4$\times$10$^4$ yrs: 
 not surprisingly the inner parts of the core show an increase of the molecular fractional abundances
for most of the species while they are very low in the outer regions of the core. Such trends are also present 
as a function of time as shown in Figures 2 and 3, where 
fractional abundances of selected species as a function of (logarithmic) time at visual 
extinction of $\sim$ 140 mags for a warm core at two different densities, are presented 
($2.0 \times 10^{8}$cm$^{-3}$, solid line, and $1.0 \times 10^{7}$cm$^{-3}$, dashed line). 

All the species in our sample show sudden increases in their abundances corresponding to 
their multi-step sublimation from grain mantles (see \citealt{col04}). 
However, the fractional abundances of H$_2$CO and CH$_3$OH are interesting because they show 
extra `jumps' which cannot be explained by sublimation from grains. 
The increase in the abundances of H$_2$CO and CH$_3$OH around these jumps is
associated with the `volcano' desorption of CH$_4$ from grains. 
The increase in H$_2$CO and CH$_3$OH seen in conjunction with the CH$_4$ `volcano' desorption is
due to a sequence of gas-phase reactions involving CH$_4$. Although, in our model, both H$_2$CO 
and CH$_3$OH are formed on grains, via the successive hydrogenation of CO 
(as confirmed both experimentally and theoretically (e.g. \citealt{tiel82, wat03}) 
there is still a contribution from 
gas-phase chemistry, especially for H$_2$CO. Our detailed chemical
analysis shows that at the time of the jump, the formation of H$_2$CO is
dominated by the oxidisation of CH$_3$ which is in turn formed efficiently by
the destruction of CH$_4$ by He$^+$ ions. Methanol is found to be a secondary
product of CH$_5^+$ which is mainly formed (at that time) via two ion-molecule
reactions involving CH$_4$; one with H$_3^+$ and the other with N$_2$H$^+$. Destruction of
CH$_5^+$ leads to CH$_3^+$ which reacts with water to form CH$_3$OH$_2^+$ which recombines
to give methanol. These sequences of 
reactions which involve CH$_4$ contribute to the observed amount of both H$_2$CO and CH$_3$OH at the `jump'. 
In fact, the `jump' in abundance for methanol is rather small because the CH$_3$OH$_2^+$ recombination 
reaction is rather inefficient (see \citealt{gep06})

For most species, the abundances are
lower at lower density with the exceptions of SO and SO$_{2}$. SO is
mainly produced by reactions involving atomic and molecular oxygen during 
early and late times respectively,  while it is destroyed by HCO$^+$ via radical-molecule 
reactions which are more pronounced during late times. SO is also destroyed 
while forming SO$_2$ via oxidisation and radical-molecule reactions, during
both early and late times. At high densities, SO is destroyed more efficiently 
than in less dense cores. 
For less dense cores, the chemistry reaches steady state earlier and
the evaporation peaks are more pronounced. 

In Figures 4 and 5 we show models with depletion efficiencies of 100\% and 85\% respectively. 
Lowering the depletion efficiency reduces the yield of most species.  However, some species
(Fig. 4; e.g. SO, SO$_2$) show higher abundances at lower depletion efficiencies: 
this is due to the higher abundance of atomic sulphur in the gas phase during Phase I 
which leads to an appreciable amount of SO and SO$_2$ even before the warm core forms. 
In general, models where all heavy elements are frozen out give abundances that are
more in line with the observed abundances (see Table 2 and Section 3.2).
We have also run a model where sulphur does not hydrogenate as it freezes but remains in atomic form 
\citep{wak04} and, as expected, we find that this affects the early stages of Phase II by lowering 
the abundance of H$_2$S by a factor of 30; however,  by 10$^5$ yrs the bulk of the sulphur is locked 
in CS, SO and SO$_2$, regardless of its initial form. 

From our results so far it is clear that, as in the case of hot cores, sulphur-bearing species show
the most variation, whether with time, radius, density or depletion. Ratios of sulphur-bearing molecules
may therefore be used as tracers of the physical characteristics of warm cores. The SO/CS, SO/H$_{2}$CS, 
and SO/OCS ratios are less than 1 at early times ($\leqslant 1 \times 10^{4}$ yrs), while those of 
H$_{2}$S/SO$_{2}$ and H$_{2}$S/H$_{2}$CS are higher than 2 at later times ($\geqslant 6 \times 10^{6}$ yrs). 
The slowing down of the temperature increase allows more species to survive in the gas-phase, and hence show higher 
abundances at late times. This result is in line with the prediction of both \citet{hat98} and \citet{viti04} 
for the change of the abundances of sulphur-bearing species. The key difference  
between low and high mass cores is the abundance of H$_2$S as a function of time: 
this species decreases much more drastically with time for low-mass cores. 
Note that H$_2$S drives the sulphur bearing chemistry (see \citealt{hat98,viti04}).
These ratios may therefore, in principle, be used as evolutionary indicators for warm 
cores, although, as \citet{wak04} point out, care needs to be taken because sulphur-bearing ratios at early 
times strongly depend on the nature of sulphur in the icy mantles.

\subsection{Comparison with hot core chemical models}
The main aim of this study was to determine whether the predicted
chemistry of warm cores differs substantially from that of hot cores and in particular whether we 
can identify molecules that would be enhanced in warm cores but not in hot cores.
\citet{viti04} ran chemical models of hot cores for different stellar masses. 
In this work we compare our results for a one solar mass warm core with those of \citet{viti04} 
for hot cores. Typical hot cores (as modelled by \citealt{viti04}) have temperatures of 
300 K, sizes of about 0.03 pc, and densities of the order of 10$^7$ cm$^{-3}$; in this paper
we concentrate on the inner part of a typical warm core (or `hot corino') 
and thus employ much smaller sizes, higher densities and lower temperatures 
(see Table \ref{tab:initial}). These differences in initial conditions lead to 
two main differences in the chemistry: firstly we find that 
in general hot cores possess higher fractional abundances, at late times 
($\geqslant 2 \times 10^{5}$ yrs), while those abundances are higher in warm cores at early stages. 
This is most likely due to the higher densities in warn cores (leading to a richer chemistry 
both in both phases) and the different evaporation times (leading to different species being 
present in the gas phase at different times). 
Secondly, the evolutionary profiles of the species show almost the same trends shifted toward later 
times for hot cores. 
The observed jump in the abundances of species (such as H$_{2}$CO) in warm cores (e.g. \citealt{cec96,cec20a,
cec20b}) is not observed for the hot core case, which may indicate that these jumps are essential features in the 
warm core evolutionary profiles for those species. 
COMs show high abundances comparable to those found in hot cores, in particular for CH$_{3}$OH, CH$_{3}$CN and 
C$_{2}$H$_{5}$OH. From the above discussion we conclude that the set of species used as evolutionary indicators 
for hot cores may be used for the same purpose in warm cores.

\subsection{Comparisons with observations of low-mass cores}
While we do not attempt here to model any particular observed warm core,
the physical parameters of one of our models with density of 2.0 $\times$ 10$^8$ cm$^{-3}$
seem to be consistent with those observed
toward the well-studied warm core around the solar type
Class 0 source IRAS 16392-2422 (e.g \citealt{cec99,cec04asp,cec07msl}). 
These authors find that at a radius of  $150$ AU and density $2.0 \times 10^{8}$ cm$^{-3}$, 
the dust temperature is $\sim 100 $ K. Hence we may 
qualitatively compare our results to the observations of this core. 
\citet{cec20a,cec20b} recorded sudden (spatially abrupt) increases in the
abundances of a few observed species such as H$_2$O, SiO, and
H$_2$CO. \citet{sch02} found that in order to reproduce the physical 
properties of the dust and gas components
constituting the material in the circumstellar envelope of IRAS 16293-2422 
a jump in the abundance of H$_2$CO and other species was indeed needed. Both of 
these papers proposed that these jumps, recorded around 80-90 K, 
result from the evaporation of mantle species in the hot-core like 
region around the protostar. Our model with density of 2.0 $\times$ 10$^{8}$ cm$^{-3}$
does indeed reproduce the
jumps in the profiles of H$_2$CO and CH$_3$OH, around 84 K, confirming their findings. 

A tentative comparison between our model calculations averaged through the core 
and observational fractional abundances in IRAS 16293-2442 is shown in Table 2, 
at time around 8.5$ \times$ 10$^{4}$ yrs, when the dust temperature reaches $\sim 100$ K. 
Because our model calculations are for a core of radius 150 AU, we compare our results 
with the observations of the inner core by \citet{sch02}, apart from the measurement of 
CH$_3$CN that was taken from \citet{bis08}.

\begin{table*}
\begin{center}
\caption{Our calculated fractional abundances in comparison with observations of IRAS 16293-2422.} 
\begin{tabular}{lllll}
 \hline
 \hline {\bf 
Species  }     & This work             & Observations\\
\hline
OCS            &$ 3-12 \times 10^{-9}$ & $  2 \times 10^{-7}$ \\
SO             &$ 1-20 \times 10^{-8}$ & $  2 \times 10^{-7}$ \\
SO$_2$         &$ 5-12 \times 10^{-8}$ & $  1 \times 10^{-7}$ \\
H$_2$CO        &$ 9-25 \times 10^{-9}$ & $  6 \times 10^{-8}$ \\
CH$_3$OH$\ddag$&$ 2-3 \times 10^{-7}$  & $  3 \times 10^{-7}$ \\
CH$_3$CN       &$ 6-13 \times 10^{-9}$ & $ 6 \times 10^{-8}$\\
\hline
\end{tabular}
\label{Table-object-dist}
\end{center}
\flushleft
$\ddag$ Our calculations for methanol are for core radius of $\leq 120$ AU, 
after that the fractional abundance drops to $\sim 10^{-12}$
\end{table*}

Sulphur-bearing species are, on average, in good agreement with observations during 
the early evolutionary stages ($t \leq 7 \times 10^{4}$ yrs) and hence they are 
good tracers for young cores with temperatures $\leq 84$ K. On the other hand, we 
find that large complex molecules show good agreement with observations at later 
times ($t \geq 9 \times 10^{4}$ yrs) where the temperature exceeds 100 K. 
This result supports the view of grain mantles as factories of complex species and 
it underlines the importance of an accurate desorption treatment in the chemical modelling 
to provide abrupt changes in molecular abundances.

\section{Conclusions}
We report here results from a chemical model of a warm core (or hot corino) 
around a low-mass star. The core contains a protostellar object of solar mass 
whose radiation warms its inner part. This leads to sublimation of icy mantles deposited
during the preceding collapse phase. We have introduced a detailed model
of sublimation based on extensive laboratory results of \citet{col04}. 
These experiments imply that sublimation from mixed ices during
the warming process occurs in several well-defined temperature bands. We
have computed the positions of these bands for the solar-mass model and
follow the resulting chemistry in a time- and space-dependent manner.

Although we have not attempted to model any specific astronomical source,
the physical parameters we have used are sufficiently similar to those
inferred from observations of a well-studied object, IRAS 16293-2422,  for us to compare
the model results with observations. The predicted and inferred molecular
abundances agree reasonably well.

The model that we have used is essentially a hot core model in which
several physical parameters have been adjusted to be appropriate to
warming by a solar-mass protostar. The generally satisfactory behaviour of
the model results suggests that the chemical processes for
both warm cores in low-mass star forming
regions and hot cores in high mass star forming regions are similar.

Observations of warm cores show the existence at particular values of A$_V$
of very substantial `jumps' in abundance (by several orders of magnitude)
in certain molecular abundances. We find that these jumps arise as a
consequence of sublimation and subsequent rapid gas-phase chemistry. Such
jumps do not occur in models that describe sublimation as an instantaneous
process, rather than the staged process we have introduced.

We predict that ratios of certain sulphur species should be excellent
tracers of physical conditions in warm cores, regardless of the (unknown)
total sulphur abundance.

\section{Acknowledgements}
Z. Awad would like to thank both the ORS and the Perren studentship 
schemes for funding. The authors thank the referee for helpful comments 
on the original version of this paper.

\newcommand{\apj}[1]{ApJ, }
\newcommand{\mnras}[1]{MNRAS, }
\newcommand{\aj}[1]{Aj, }
\newcommand{\apjs}[1]{ApJS, }
\newcommand{\apjl}[1]{ApJ Letter, }
\newcommand{\aap}[1]{A\&A, }
\newcommand{\aaps}[1]{A\&A Suppl. Series, }
\newcommand{\araa}[1]{Annu. Rev. A\&A, }
\newcommand{\aaas}[1]{A\&AS, }
\newcommand{\apss}[1]{Ap\&SS }
\newcommand{\bain}[1]{Bul. of the Astron. Inst. of the Netherland,}

\bibliographystyle{mn2e}


\begin{figure*}
\centering
\includegraphics[width=15cm]{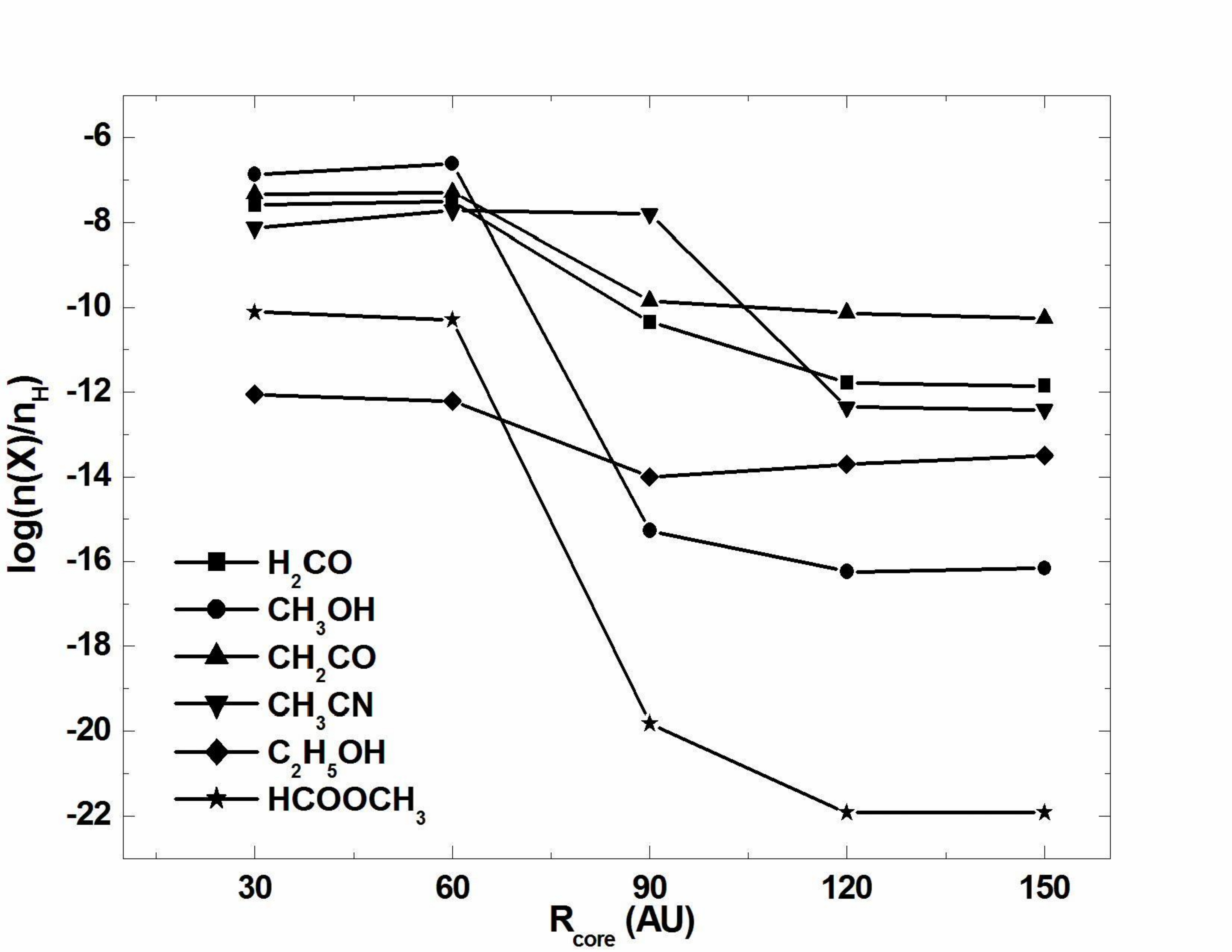}
\caption{Fractional abundances with respect to the total hydrogen 
nuclei as a function of the core radius (R$_{core}$) for selected organic 
species at 4$\times$10$^4$ years for a model with density of 2$\times$10$^{8}$ cm$^{-3}$.}
\label{fig:Fig1}
\end{figure*}

\begin{figure*}
\centering
\includegraphics[width=20cm]{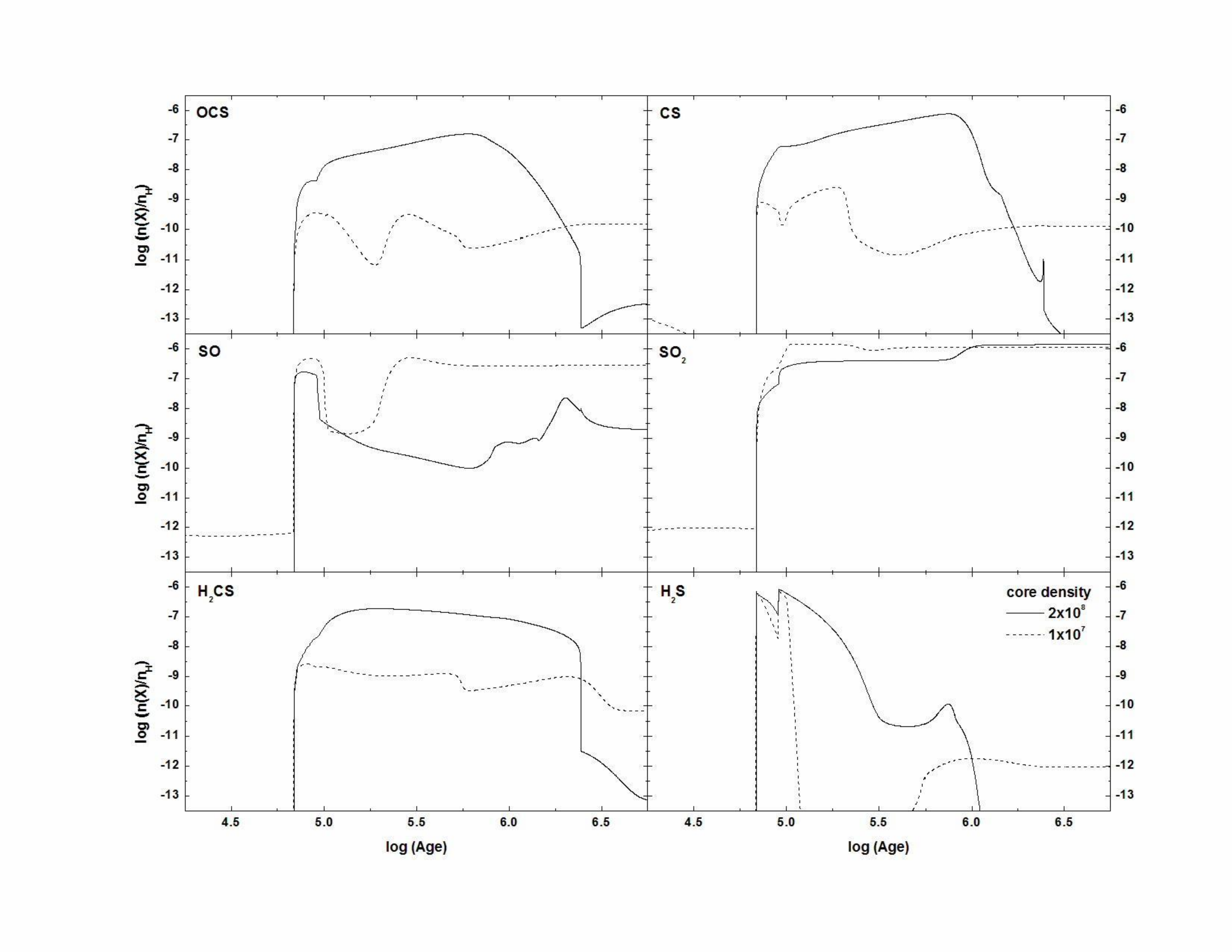}
\caption{The time evolution of the fractional abundances of S-bearing 
molecules at $A_v = 142.7$ mag. 
The different curves compare the evolution of the species at two different 
final densities for the collapsing cloud (see key).}
\label{fig:fig2}
\end{figure*}

\begin{figure*}
\centering
\includegraphics[width=20cm]{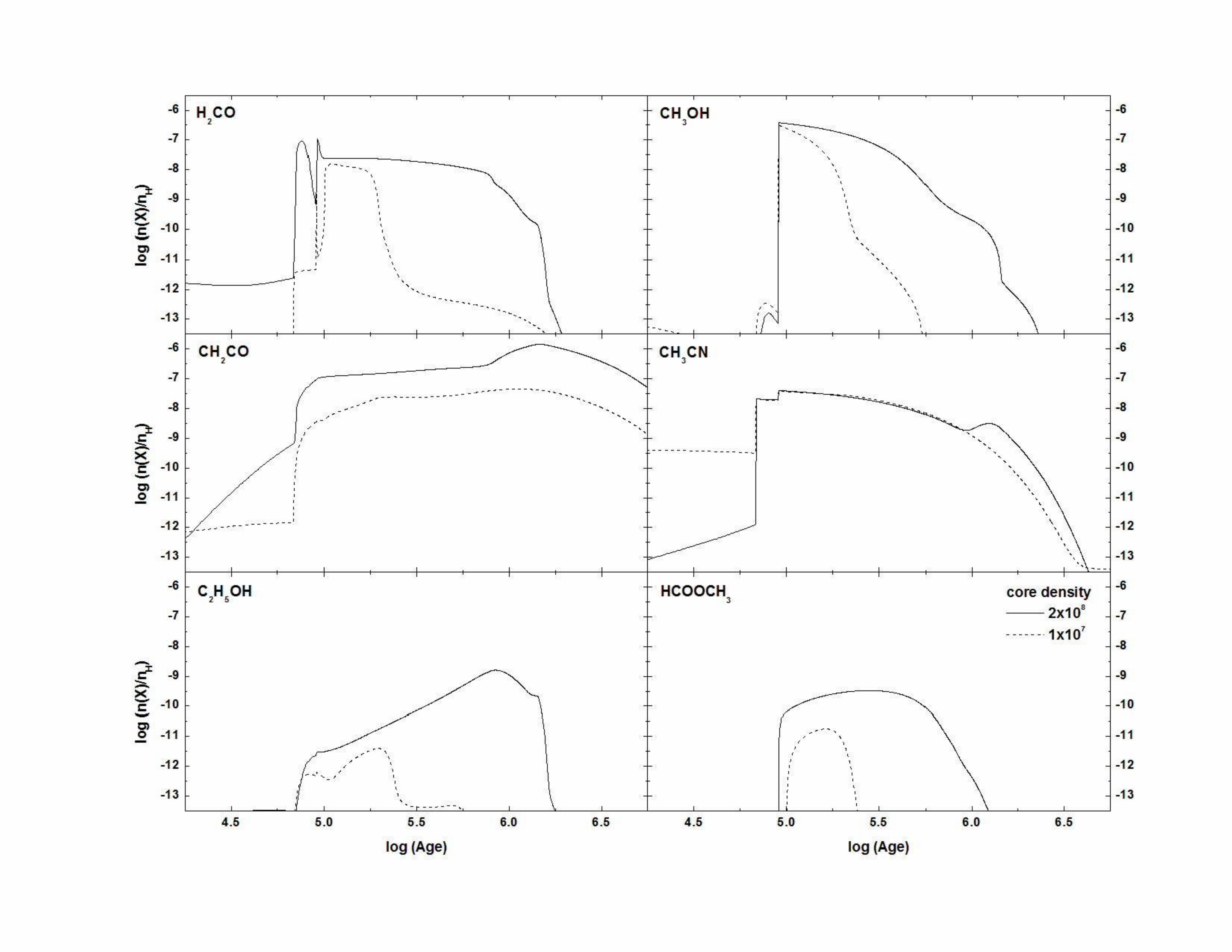}
\caption{The same as in Fig. (2), but for large and organic molecules.} 
\label{fig:fig3}
\end{figure*}

\begin{figure*}
\centering
\includegraphics[width=20cm]{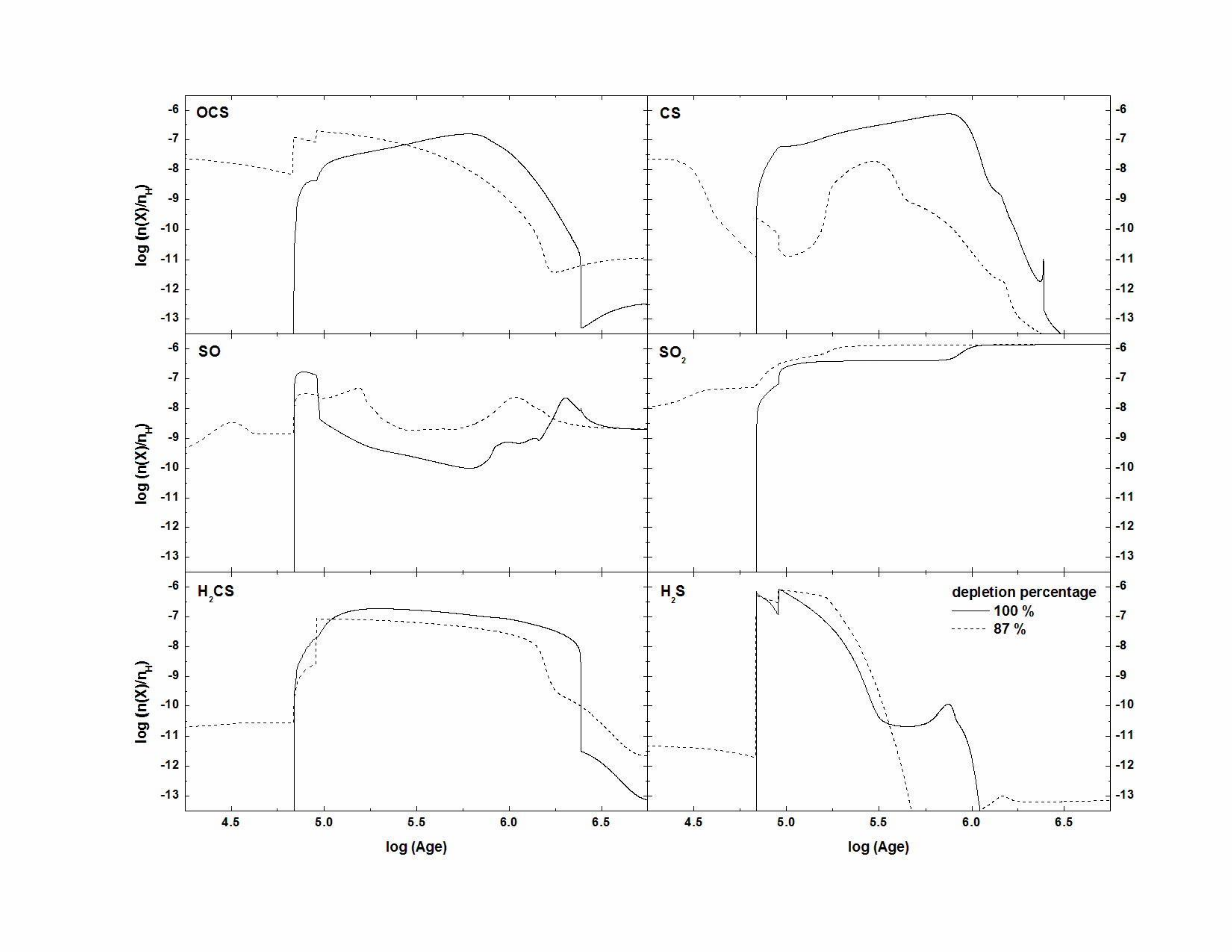}
\caption{The time evolution of the fractional abundances of 
S-bearing molecules in a warm core region, of density 
$n_H= 2.0\times10^8$ cm$^{-3}$, at a point with $A_v = 142.7$ mag.
The profiles compare the evolution of the fractional abundances at 
different depletions on the grain surfaces (see key).}
\label{fig:fig4}
\end{figure*}

\begin{figure*}
\centering
\includegraphics[width=20cm]{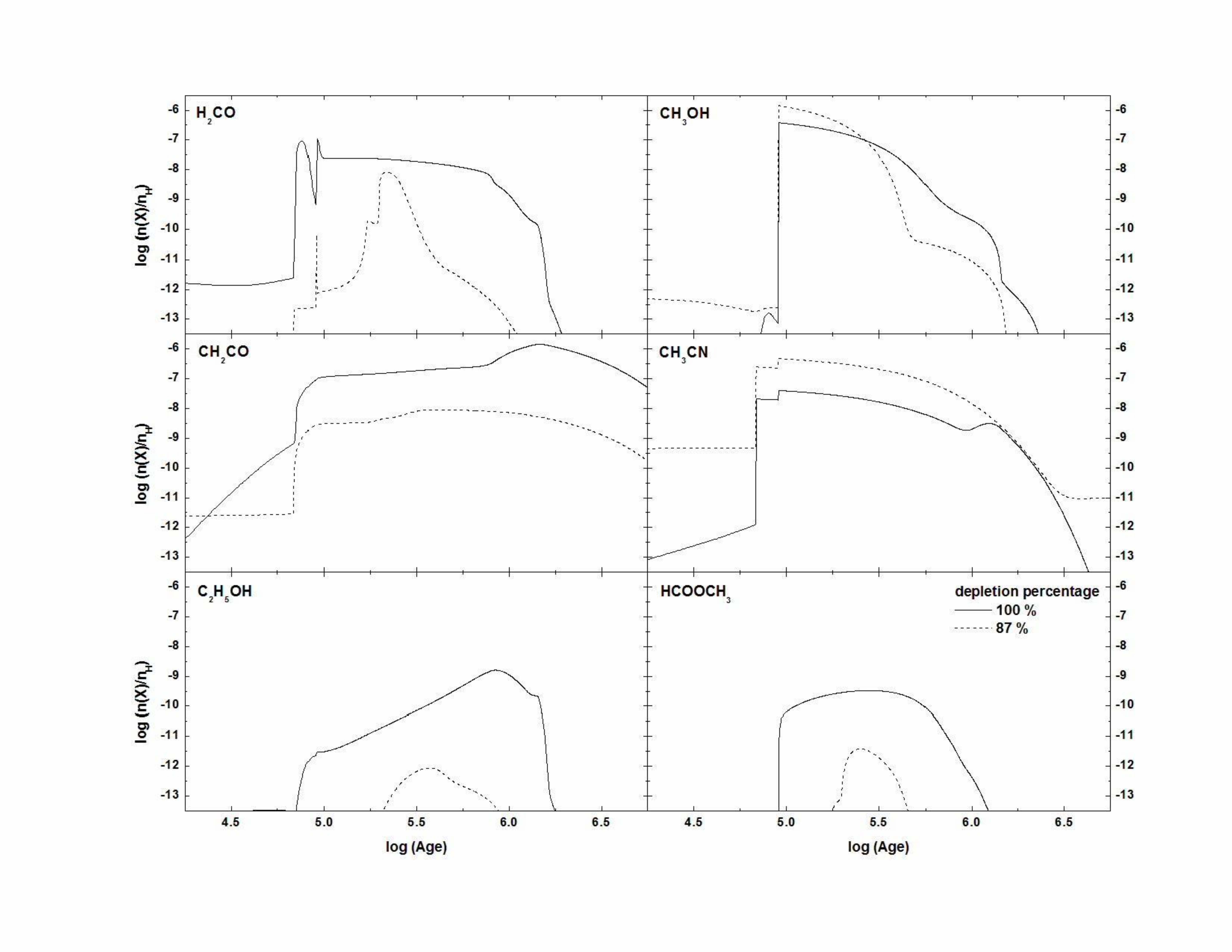}
\caption{As in Figure (4) but for the large organic molecules}
\label{fig:fig5}
\end{figure*}

\label{lastpage}

\end{document}